\documentclass[%
reprint, 
 amsmath,amssymb,
 aps,
prb,
citeautoscript,
]{revtex4-1} 

\usepackage{graphicx}
\usepackage{dcolumn}
\usepackage{bm}

\renewcommand{\d}[1]{\ensuremath{\operatorname{d}\!{#1}}}
\DeclareRobustCommand{\matr}[1]{\bm{#1}}
\renewcommand{\vec}[1]{\bm{#1}}
\renewcommand{\imath}{i}
\DeclareMathOperator{\sgn}{sgn}

\begin{document}

\preprint{APS/123-QED}

\title{Propagation of acoustic edge waves in graphene under quantum Hall effect}

\author{A. Vikstr\"{o}m}
 \email{anton.vikstrom@chalmers.se}
\affiliation{%
Department of Applied Physics, Chalmers University of Technology,\\ Kemig\aa{}rden 1,
412 96 G\"{o}teborg, Sweden}%




\date{\today}

\begin{abstract}
We consider a graphene sheet with a zigzag edge subject to a perpendicular magnetic field and investigate the propagation of in-plane acoustic edge waves under the influence of magnetically induced electronic edge states. In particular is is shown that propagation is significantly blocked for certain frequencies defined by the resonant absorption due to electronic-acoustic interaction. We suggest that strong interaction between the acoustic and electronic edge states in graphene may generate significant non-linear effects leading to the existence of acoustic solitons in such systems.
\end{abstract}

\maketitle




The discovery of graphene\cite{Novoselov22102004}, an ultra-pure 2D crystal membrane of remarkable promise\cite{RiseOfGraphene}, has in just the past few years led to the rapid growth of a new field of research, uniting and challenging scientists from research backgrounds as diverse as the capabilities of the material itself. In addition to its astounding material properties, the very existence of a true 2D crystal both requires and inspires new ways of thinking.

It is well known that a 3D continuous medium supports acoustic waves localized to the surface\cite{ToE}. Such surface waves have been used to probe the electronic properties of samples\cite{0022-3719-17-13-024}, e.g. the fractional quantum Hall effect (QHE) of 2D electron gasses in semi-conductor heterostructures\cite{PhysRevLett.65.112,PhysRevB.47.7344}, topological insulators\cite{PhysRevB.83.125314,PhysRevB.54.13878} and, more recently, graphene\cite{PhysRevB.81.041409}. In past schemes the surface wave direction of localization was normal to the 2D electron gas plane so that the electrons experienced no localization of acoustic energy. However, the isolation of single-layer graphene\cite{Novoselov22102004}, a flexible 2D membrane, suggests the existence of acoustic \emph{edge waves}, a 2D analog of the 3D surface waves. Recent studies have shown such edge-localized vibrational motion in graphene to consist of both in-plane and flexural modes, both which decay into the 2D ``bulk''\cite{PhysRevB.81.165418}. At the same time, a magnetic field applied perpendicularily to the sheet would induce current-carrying electronic states localized to the same graphene edge on the order of the magnetic length, $l_B=\sqrt{\hbar / | e B |} \approx 26 \text{ nm}/\sqrt{B[T]}$ ($B[T]$ is the dimensionless field strength in Tesla)\cite{PhysRevB.73.125411,gusynin:778,PhysRevB.77.205409,PhysRevB.79.115431,PhysRevB.83.045421,RevModPhys.81.109,PhysRevB.73.195408}. In this paper we investigate the interaction between electronic quantum Hall effect edge states and localized acoustic edge waves, specifically low-amplitude in-plane \emph{Rayleigh waves}\cite{ToE}, while flexural modes will be neglected.

\begin{figure}
  \centering
  \includegraphics[width = \columnwidth]{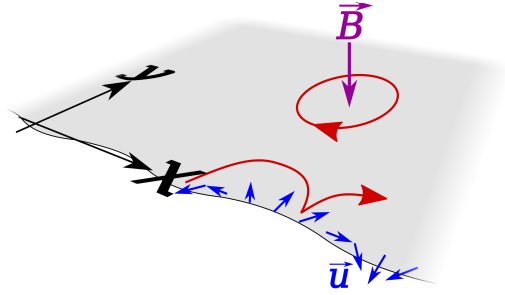}
  \caption{(Color online) A schematic picture of a continuous (graphene) sheet with an edge along the $x$-axis and an applied perpendicular magnetic field (purple). The electronic states (red) may be either localized Landau orbits in the bulk or dispersive states near the edge. Along the edge there are propagating acoustic (Rayleigh) edge waves given by a 2D displacement field (blue, amplitude exaggerated).}
  \label{fig:geometry}
\end{figure}

To be concrete, we consider a 2D graphene sheet with a stress-free zigzag edge at $y=0$, directed along the $x$-axis, see Fig. \ref{fig:geometry}. A transverse magnetic field, $\vec{B}=-|B| \vec{e}_z$, is then applied to the sheet ($\vec{e}_{x,y,z}$ are unit vectors), bringing the sample into the quantum Hall effect regime. The sheet is treated as a continuous medium and the width of the sample is taken to be large enough for the electronic and acoustic edge states to decay completely across the the sample; it is then enough to consider only one edge. The sample length $L$ is assumed to be long enough to allow for acoustic wave propagation in the $x$-direction.

Since the graphene edge, which is normal to $- \vec{e}_y$ and located at $y=0$, is stress-free, the elastic boundary conditions are
\begin{equation}
\sigma_{j y}(x,0)=0 , \; \; j=x,y,z \; , \label{eq:ToEbc}
\end{equation}
where $\sigma_{ij}(x,y)$ is the usual 2D stress tensor\cite{ToE}. Since the Rayleigh waves are pseudo-1D, they can be specified by the wave vector $x$-component $q$ alone, which will be referred to as the \emph{wave number}. Standard techniques\cite{ToE} give the two-component displacement field $\vec{u}^{(q)}(x,y)$ for an in-plane Rayleigh wave as
\begin{equation}
	\vec{u}^{(q)}(x,y;t) = 2 u_0 \begin{pmatrix}
 f^{(q)}_x (y) \cos (qx-\omega t ) \\ 
 \sgn (q) f^{(q)}_y (y) \sin (qx-\omega t )
\end{pmatrix} \label{eq:realdispl} \; , 
\end{equation}
where
\begin{equation}
\vec{f}^{(q)}_x (y) =  {e}^{- \lambda_l | q | y} - C_x {e}^{-\lambda_t | q | y}
\label{eq:f_x}
\end{equation}
and
\begin{equation}
\vec{f}^{(q)}_y (y) = - \lambda_l {e}^{- \lambda_l | q | y} + C_y {e}^{-\lambda_t | q | y} .
  \label{eq:f_y}
\end{equation}
The dimensionless constants are
\begin{align}
	\lambda_l = 0.81 , \; \;
	\lambda_t = 0.46 , \nonumber \\
	C_x = 0.61  , \; \;
	C_y = 1.3 \; ,
\end{align}	
and depend only on the ratio of the transverse and longitudinal sound velocities in graphene, $s_t/s_l$, or, equivalently, on the Poisson ratio. The sound velocities are taken to be $s_t=1.4 \cdot 10^4 \text{ m/s}$ and $s_l=2.1 \cdot 10^4 \text{ m/s}$\cite{PhysRevLett.105.266601,PhysRevB.85.165440}. The dispersion relation is linear,
\begin{equation}
	\omega{(q)} = s_R |q|	\label{eq:omega} \; ,
\end{equation}
with Rayleigh-wave sound velocity $s_R=1.2 \cdot 10^4 \text{ m/s}$.


The electronic subsystem is described by the standard effective-model graphene Hamiltonian
\begin{equation}
\hat{\matr{H}}_{\text{el}}=v^F (\matr{\sigma}_x \hat{p_x} + \tau \matr{\sigma}_y \hat{p_y}) ,
\label{eq:H_el}
\end{equation}
where $v^F = 1.0 \cdot 10^6 \text{ m/s}$ is the Fermi velocity of graphene, $\tau = +1$ ($-1$) for the $K$-point ($K'$-point), the $\matr{\sigma}$s are the sublattice-space Pauli matrices\cite{RevModPhys.81.109,PhysRevB.54.17954,PhysRevB.73.195408} and the sublattice psuedospinor upon which the Hamiltonian acts is defined by $\vec{\psi}(x,y)=(\psi_A(x,y),\tau \psi_B(x,y))^T$. The transverse magnetic field is represented by a vector potential in the Landau gauge, $\vec{A}_B=(By,0)^T$, and then included in the Hamiltonian of Eq. (\ref{eq:H_el}) through the minimal coupling $\vec{p} \rightarrow \vec{p} + e \vec{A}$ (the electron charge is $-e<0$). 

In an infinite bulk system the electronic energies form Landau levels\cite{Natv438,PhysRevLett.98.197403,NatPhysv3i9},
\begin{align}
	E_n &= \sgn (n) E_1 \sqrt{| n |} , \; \; n= 0, \pm 1, \pm 2, \ldots \nonumber \\
	E_1 &= \sqrt{2} \hbar v^F l_B^{-1} \propto \sqrt{|B|} \; ,
	\label{eq:LL}
\end{align}
and the electronic wave~functions are harmonic oscillator states centered around $y_c=-k l^2_B$, corresponding to closed Landau orbits, see Fig. \ref{fig:geometry}. This simple picture is modified by the introduction of an edge. 

In the considered system the edge at $y=0$ is a zigzag edge of $B$-atoms, leading to the electronic boundary condition\cite{PhysRevB.73.235411} 
\begin{equation}
  \psi_A(x,0)=0 .
  \label{eq:elbc}
\end{equation}
Since the zigzag boundary condition does not mix valleys the $K$- and $K'$-points can be considered separately. 

The edge induces a positive (negative) dispersion in the electron-like (hole-like) Landau levels as $k$ increases and the wave~function center $y_c \propto -k$ moves toward and over the edge\cite{PhysRevB.73.125411}, pressing the oscillator wave~functions against the edge and turning them into edge-localized current-carrying states. For a classical, intuitive picture of this effect, see Fig. \ref{fig:geometry}. 

\begin{figure}
  \centering
  \includegraphics[width = \columnwidth]{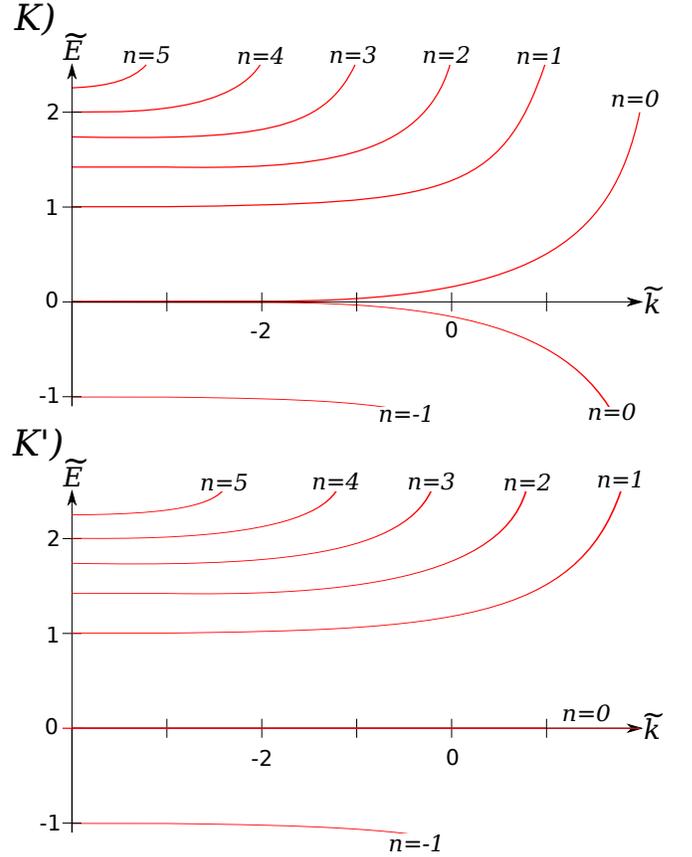}
  \caption{(Color online) A schematic picture of the electronic spectrum around the points $K$ (top) and $K'$ (bottom). The scaled energy $\tilde{E}=E/E_1$ is plotted against the scaled wave number $\tilde{k}=k l_B$ (energy bands in red). The leftmost low-$\tilde{k}$ states are bulk states and their spectrum consists of discrete Landau levels. The dispersive states are edge states and here share the label $n$ with their bulk counterparts.}
  \label{fig:spectrum}
\end{figure}

The dispersion can be calculated by generalizing the Landau-level~index $n$ to a \emph{continuous} analogue, $\nu=(E/E_1)^2$, and the harmonic oscillator functions to (Whittaker's) parabolic cylinder functions $D_{\nu}(z)$, which reduce to harmonic oscillator functions for integer $\nu$ but allow for non-integer $\nu$ solutions between the bulk Landau levels. The spectrum is then calculated from the boundary condition of Eq. (\ref{eq:elbc})\cite{gusynin:778,PhysRevB.77.205409,PhysRevB.79.115431,PhysRevB.83.045421}. The dimensionless energy $E/E_1 \equiv \tilde{E}$ is plotted against the dimensionless wave number $k l_B \equiv \tilde{k}$ in Fig. \ref{fig:spectrum} for both the $K$- and $K'$-points. 
The energy band stemming from Landau level $n$ will be referred to as ``edge band $n$''. When $\tilde{k}=k l_B=0$, $y_c=0$ and the wave~function is centered on the edge.

As seen in Fig. \ref{fig:spectrum}, the zeroth Landau level remains dispersonless for all $\tilde{k}$ in the $K'$-point spectrum, whereas it is seemingly split in two edge bands, one electron-like and one hole-like, in the $K$-point spectrum. This can be explained by extra degeneracies introduced by topological edge states; the peculiar nature of the $n=0$ Landau level  have been studied in other papers\cite{JPSJ.65.1920,PhysRevB.78.205401,PhysRevB.79.115431,PhysRevB.83.045421}; for the purpose of this paper the schematic spectra in Fig. \ref{fig:spectrum} will suffice.

The electronic pseudospinor wave~functions are given in the appendix for reference. There, scaled physical coordinates $\tilde{x} (\tilde{y}) \equiv x (y) / l_B$ are introduced, which will be employed below when considering the absorption.

The standard first-order-in-strain Hamiltonian for the electron-strain interaction in graphene is given by \cite{PhysRevB.65.235412}
\begin{multline}
\matr{H}_{\text{int}}^{\tau} \left( \vec{u} (x,y;t) \right) =  g_1 (u_{xx} + u_{yy}) \matr{I} + \\  + g_2 \left( - \tau (u_{xx}-u_{yy}) \matr{\sigma}_x + 2 u_{xy} \matr{\sigma}_y \right) ,
\label{eq:Hint}
\end{multline}
where $u_ij$ is the standard strain tensor. The diagonal elements are the scalar deformation potential, with coupling constant $g_1\sim 10 \text{ eV}$, and the off-diagonal elements are usually imagined as a strain-induced pseudo vector-potential, and their coupling constant is $g_2 \sim 1 \text{ eV}$. Since the valley separation is $|\vec{K}-\vec{K}'| \sim a^{-1}$, interaction with the acoustic Rayleigh waves will not mix $K$ and $K'$ if the acoustic wave number $q \ll a^{-1}$, which must hold for the continuous-media model to be valid. Therefore  all electronic transitions induced by the acoustic waves are intra-valley and the $K$- and $K'$-point spectra can still be considered separately using $\tau= \pm 1 $.

Inserting Eqs. (\ref{eq:realdispl}), (\ref{eq:f_x}), and (\ref{eq:f_y}) into Eq. (\ref{eq:Hint}) yields the Hamiltonian for an electronic transition due to interaction with the acoustic Rayleigh waves as
\begin{multline}
  \matr{H}_{\text{int}}^{\tau} \left( \vec{u}^{(q)}(x,y;t) \right) = u_0 e^{\imath q x - \imath \omega t} ( \imath q ) \left\lbrace g_1 (T_1 \matr{I} e^{- \lambda_l |q| y} + \right. \\
   \left. + g_2 \left( \left[ - \tau T_2^{x,l} \matr{\sigma}_x + \imath \sgn (q) T_2^y \matr{\sigma}_y \right] e^{- \lambda_l |q| y} + \right. \right. \\
   \left. \left. + \left[ \tau T_2^{x,t} \matr{\sigma}_x - \imath \sgn (q) T_2^y \matr{\sigma}_y \right] e^{- \lambda_t |q| y} \right)  \right\rbrace + \text{H.c.}
\label{eq:finHint}
\end{multline}
where the constants are
\begin{align}
	T_1 = 0.34 , \; \;
	T_2^y = 1.6 , \nonumber \\
	T_2^{x,l} = 1.7  , \; \;
	T_2^{x,t} = 1.2 \; .
\end{align}


Considering the spectra for the $K$- and $K'$-points in Fig. \ref{fig:spectrum}, it is evident that using a gate voltage $V_G$ to adjust the scaled Fermi energy, $E^F/E_1=\tilde{E}^F \propto V_G / \sqrt{B}$, alters the number of dispersive energy bands crossing the Fermi level. If 
\begin{equation}
 | \tilde{E}_{n-1} | <  | \tilde{E}^F | < | \tilde{E}_{n} | , \label{eq:fermireq}
\end{equation}
where $\tilde{E}_n$ refers to the scaled energy of bulk Landau level $n$ (see Eq. (\ref{eq:LL})), there will be $n$ ($n-1$) energy bands crossing the Fermi level in the $K$-spectrum ($K'$-spectrum). These crossings are the quantized conduction channels of the quantum Hall effect theory and the absolute values in Eq. (\ref{eq:fermireq}) correspond to the electron-hole symmetry of the spectrum. The dispersionless level in the $K'$-spectrum never crosses the Fermi level and is therefore assumed never to be involved in transitions. 

To analyze the possible transitions, consider the transition rate between levels, thereby introducing conservation laws. The transition rate $W_{m,n}$ for an electronic jump from energy band $n$ to energy band $m$ due to interaction with an acoustic wave with scaled wave number $q l_B \equiv \tilde{q}$ is given by the Fermi golden rule, 
\begin{multline}
  W_{m,n} = \frac{2 \pi}{\hbar} \sum_{\tilde{k}_n} \int \d{\tilde{E}_{m}} \delta \left( \tilde{E}_n + \tilde{E_R} - \tilde{E}_{m} \right)  \delta_{\tilde{k}_n + \tilde{q},\tilde{k}_m} \times \\ 
  \times \rho ( E_m ) \left| \Lambda_{\tilde{k}_m;\tilde{q};\tilde{k}_n}^{\tau} \right|^2  
   f_{\text{FD}} (E_n) \left( 1 - f_{\text{FD}} (E_m) \right) . \label{eq:fgr}  
\end{multline}
Here, $\delta_{\tilde{k}_n + \tilde{q},\tilde{k}_m} \Lambda_{\tilde{k}_m;q;\tilde{k}_n}^{\tau}$ is the matrix element of an induced transition from $\tilde{k}_n$ to $\tilde{k}_m$ defined by
\begin{multline}
 \delta_{\tilde{k}_n + \tilde{q},\tilde{k}_m} \Lambda_{\tilde{k}_m;\tilde{q};\tilde{k}_n}^{\tau} = \\
l_B^2 \iint_\mathit{S} \vec{\psi}^{\tau,\tilde{k}_m \dagger}_{\nu} \matr{H}_{\text{int}}^{\tau} \left( \vec{u}^{(q)} \right) \vec{\psi}^{\tau,\tilde{k}_n}_{\nu} \d{\tilde{x}} d{\tilde{y}} \label{eq:lambda}
\end{multline}
where the interaction is given by Eq. (\ref{eq:finHint}) (the harmonic time dependence is accounted for by the energy conservation), the electronic wave~functions are given in the appendix and the integration surface is the whole sheet $\mathit{S}$ in terms of the scaled coordinates. The continuous level index is $\nu=(E/E_1)^2$ as before, $f_{\text{FD}} (E (k))$ is the Fermi-Dirac distribution function, $\rho ( E_m )$ is the density of final states, $\tilde{k}_n$ is the scaled wave number for an electronic state in energy band $n$ corresponding to energy $\tilde{E}_n$, and the scaled acoustic dispersion is given by, using Eq. (\ref{eq:omega}),
\begin{equation}
  \tilde{E}_R \left( \tilde{q} \right) =  \tilde{s}_R | \tilde{q} | \; , \label{eq:acEtilde}
\end{equation}
with dimensionless speed of sound
\begin{equation}
  \tilde{s}_R \equiv  \frac{s_R}{\sqrt{2} v^F} \; . \label{eq:tildeS}
\end{equation}

\begin{figure}
  \centering
  \includegraphics[width = \columnwidth]{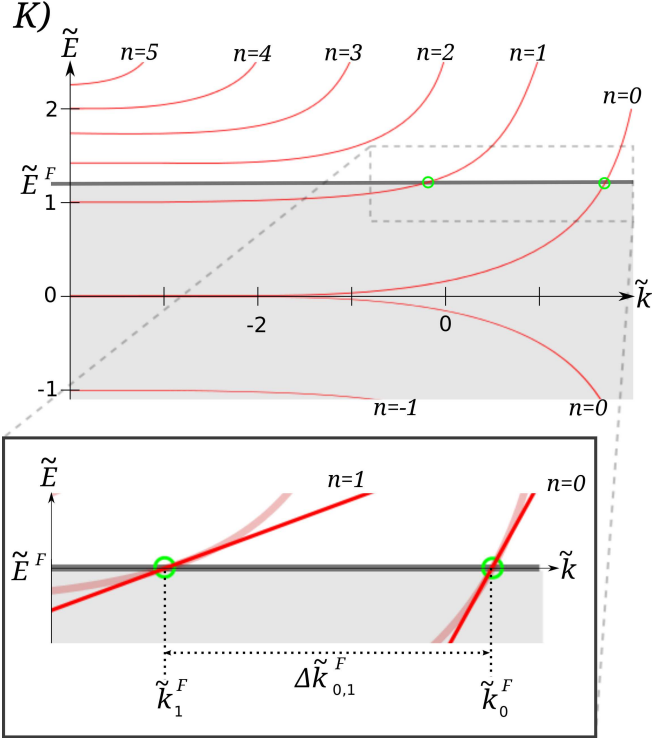}
  \caption{(Color online) In the $K$-point spectrum of Fig. \ref{fig:spectrum} the Fermi energy $\tilde{E}^F$ (horizontal grey line) is set by a gate voltage to lie between, say, bulk Landau level 1 and Landau level 2, thus giving the spectrum two Fermi crossing points (green circles), at $\tilde{k}^F_1$ and $\tilde{k}^F_0$, for energy band $1$ and $0$ respectively. Since transitions occur only near the Fermi level, the spectrum can be linearized, resulting in an effective model with two \emph{linear} bands crossing the Fermi level at points $\tilde{k}^F_1$ and $\tilde{k}^F_0$ (see magnified inset). The resonant frequency is then given by the wave number separation at the Fermi level $\Delta \tilde{k}^F_{0,1}=|\tilde{k}^F_0 - \tilde{k}^F_1|$. The picture is schematic.}.
  \label{fig:linspec}
\end{figure}

The energy integration and the Fermi-Dirac factors confine the energy region of absorption to the vicinity of the Fermi energy, $E_n \lesssim E^F \lesssim E_m$, and thus imply that the energies and wave numbers may be taken at the Fermi level, e.g. $\tilde{k}_n \rightarrow \tilde{k}^F_n$. Armed with this knowledge, the picture can be simplified by linearizing the spectrum, swapping each curved energy band $n$ for a \emph{linear} band $n$ with velocity equal to the Fermi velocity $v_n$ of the band, see Fig. \ref{fig:linspec}. Then the linearized dimensionless dispersion of band $n$ is
\begin{equation}
	\tilde{E}_n \left( \tilde{k}_n \right) = \tilde{v}_n \left( \tilde{k}_n - \tilde{k}^F_n \right) + \tilde{E}^F
	\label{eq:dimlesslinE}
\end{equation}
where the dimensionless velocity of the band is defined analogously to Eq. (\ref{eq:tildeS}),
\begin{equation}
  \tilde{v}_n \equiv \frac{ v_n }{ \sqrt{2} v^F } , \label{eq:dimlv}
\end{equation}
and $\tilde{s}_R \ll \tilde{v}_n \forall n$.

The above arguments together with energy and momentum conservation restrict the number of allowed transitions by imposing the requirement that
\begin{equation}
	\tilde{q} \approx \tilde{k}^F_m - \tilde{k}^F_n \equiv \Delta \tilde{k}^F_{m,n} ,
	\label{eq:DeltaKdef}
\end{equation}
i.e. the acoustic wave number $\tilde{q}$ must roughly match the $\tilde{k}$-separation of the two Fermi crossing points. Transitions occur \emph{in the vicinity of} the Fermi level, not \emph{at} the Fermi level, but for the purpose of this paper it is sufficient to take $\tilde{q} = \Delta \tilde{k}^F_{m,n}$. The same above arguments also imply that there are no allowed intra-level transitions, $n \neq m$.

The number of band-to-band transitions $N_t (n)$ for $n$ Fermi level crossings is then
\begin{equation}
	N_t \left( n \right) =
	\begin{cases} 
		\frac{n!}{2 (n-2)!} &\mbox{if } n \geq 2 , \\ 
		0 & \mbox{if } n < 2 \; ,
	\end{cases}
	\label{eq:No.trans}
\end{equation}
and it must be remembered that transitions can occur in both the $K$- and $K'$-spectra.

Since the spacing $\Delta \tilde{k}^F_{n+1,n}$ between neighboring Fermi crossings is approximately equal for the same energy, i.e. $\Delta \tilde{k}^F_{n,n+j} \approx j \cdot \Delta \tilde{k}^F_{m,m+1}$, it is potentially useful to group the transitions in terms of how many bands they jump, i.e. a jump from band $n$ to band $n-j$ is a $j$-jump (the minus sign is due to Fermi crossings of higher-$n$ bands having larger $\tilde{k}$). For the situation with $n$ Fermi crossings in one of the valley spectra, the number of $j$-jumps is
\begin{equation}
	N_{t,j} \left( n \right) =
	\begin{cases} 
		n-j &\mbox{if } n > 2 , \\ 
		0 & \mbox{if } n \leq 2 . 
	\end{cases}
	\label{eq:No.trans_j}
\end{equation}
Summing $N_{t,j}$ for all $j<n$ yields the total number of transitions in the spectrum, $N_{t}$. Since all $j$-jumps have \emph{approximately} equal $\Delta \tilde{k}^F_{m,n}$, i.e. absorbed acoustic frequency, they might appear as a multi-peak in the absorption spectrum: $N_{t,j}$ peaks close together.

The absorbed acoustic frequencies $s l_B^{-1} \Delta \tilde{k}^F_{m,n}$ can be found by using the electronic boundary conditions to find the Fermi level crossings $\tilde{k}^F_n$, see the appendix. These frequencies are on the order of $s l_B^{-1} \sim \sqrt{B[T]} \cdot 10^{11} \text{s}^{-1}$ and depend only on the scaled Fermi energy $\tilde{E_F}$, i.e. the relative position of the Fermi level. The periods of these acoustic frequencies must be much shorter than the acoustic decay~time due to interaction with the electronic subsystem for the Fermi golden rule to remain valid.

For the linearized spectrum, Eq. (\ref{eq:dimlesslinE}), standard periodic boundary conditions in the $x$-direction yields the density of final states per unit length $\rho ( E_m )$ as
\begin{equation}
	\rho ( E_m ) = \frac{1}{2 \sqrt{2} \pi \hbar v^F \tilde{v}_m} .
	\label{eq:linDoS}
\end{equation}
As seen in Fig. \ref{fig:spectrum}, the density of states increases with edge localization, i.e. increasing $\tilde{k}$.

Since transitions occur near the Fermi level, the matrix element of transition in Eq. (\ref{eq:lambda}) is evaluated for $\tilde{q} = \Delta \tilde{k}_{m,n}$ and $\tilde{E}_n = \tilde{E}_m = \tilde{E}^F$, and is then
\begin{equation}
  \Lambda_{\tilde{k}^F_m; \tilde{k}^F_n}^{\tau} = 
\imath \Delta \tilde{k}^F_{m,n}
\left( \frac{u_0}{l_B} \right)
\left( g_1 F_{1} + g_2 F_{2} \right)
\label{eq:finLambda}
\end{equation}
where the dimensionless transition-dependent integrals have been separated into a scalar potential contribution $F_{1}$ and pseudo-magnetic-field contribution $F_{2}$; both given in the appendix. Normalization of the electronic wave~functions causes these integrals to be at the most unity.

Inserting the above into Eq. (\ref{eq:fgr}), the final expression for the absorption rate per unit length is
\begin{multline}
W_{m,n}  =  \\ 
\left( \frac{1}{\sqrt{2} \hbar^2 v^F} \right) 
\frac{ \left( \Delta \tilde{k}^F_{m,n} \right)^2 }{\tilde{v}_m } %
\left( \frac{u_0}{l_B} \right)^2
\left| g_1 F_{1} + g_2 F_{2} \right|^2 .
\label{eq:finalW}
\end{multline}
where all relevant depencies have been included explicitly for clarity. The first factor $\approx 1.6 \cdot 10^{24} \text{eV}^{-2} \text{s}^{-1} \text{m}^{-1}$ and consists of general constants, and the second factor constists of parameters specific to the transition in question and is $\sim 2$. The third is the amplitude dependence, with the amplitude scaled by the magnetic length. By assumptiom, the amplitude is low, $A \ll l_b$, causing this factor to be very small. The final factor is the coupling coefficients and the transition integrals, which are less than one by normalization, meaning that the order of magnitude is set by the coupling. Inserting the definition of the magnetic length yields $W_{m,n} \propto B$. This direct proportionality to the field comes from the $\imath q$-factor in the strain tensor and the fact that absorption occurs only for the phonon wave numbers $q$ which match the electro-magnetic spectrum and are thus are on the order of inverse magnetic length.

The total energy of the acoustic wave is\cite{ezawa}
\begin{equation}
  E_{\text{ac}}= \rho_{\text{gr}} \omega( q )^2  \iint_\mathit{S} | \vec{u}^{(q) } (x,y;0) |^2 \d{x} d{y}
 \label{eq:E_ac}
\end{equation}
where $\rho_{\text{gr}} = 7.6 \cdot 10^{-7} \text{kg}/\text{m}^2$ is the surface mass density of graphene\cite{PhysRevB.85.165440}. In this case
\begin{equation}
  \iint_\mathit{S} | \vec{u}^{(q) } (x,y;0) |^2 \d{x} d{y} = \frac{2 L u_0^2}{|q| N_{\text{ac}}^2}  \label{eq:uint} \; ,
\end{equation}
and it can be shown that
\begin{equation}
  N_{\text{ac}} = 1.2 ,
\end{equation}
whereas the energy lost to each electronic transition is simply $\hbar \omega ( q )$. The acoustic inverse decay time $\tau_D$ due to interaction with the electronic subsystem is then given by
\begin{align}
  \frac{1}{\tau_D} &= \left( \frac{N_{\text{ac}}^2 \left( \Delta \tilde{k}^F_{m,n} \right)^2}{2 \sqrt{2} \hbar v^F \tilde{v}_m \rho_{\text{gr}} l_B^2 s_R} \right) \left| g_1 F_{1} + g_2 F_{2} \right|^2 \nonumber \\
  &= \frac{2.0 \cdot 10^7 B[T]}{\text{s eV}^2} \frac{ \left( \Delta \tilde{k}^F_{m,n} \right)^2 }{\tilde{v}_m } \left| g_1 F_{1} + g_2 F_{2} \right|^2 . \label{eq:invdecay}
\end{align}

As an example, consider the simplest case. The gate voltage is adjusted in relation the magnetic field so that
\begin{equation}
  \tilde{E}^F = \frac{\tilde{E}_{1} + \tilde{E}_{2}}{2}  ,
\end{equation}
i.e. the Fermi level is now in the middle of the gap between Landau level 1 and 2. According to Eq. (\ref{eq:fermireq}) there will be $2$ bands crossing the Fermi level in the $K$-point spectrum  ($1$ in the $K'$-spectrum) and by Eq. (\ref{eq:No.trans}) there will, trivially, be $1$ possible transition ($0$ possible transitions). Eq. (\ref{eq:No.trans_j}) specifies that this one transition will be between neighboring levels. Solving Eq. (\ref{eq:Kbc}) numerically returns $\tilde{k}^F_1=-1.29$ and $\tilde{k}^F_0=0.36$, the points where the bands intersect the Fermi level. This leads to $\Delta \tilde{k}^F_{0,1}=1.65$, which will be the acoustic wave number absorbed in the transition from edge band $1$ to edge band $0$. The set $\tilde{E}^F$ means that the generalized level index is, according to Eq. (\ref{eq:nuF}), $\nu^F=((1+ \sqrt{2} )/2)^2 \approx 1.4571$ and the velocity of the destination band is estimated to $\tilde{v}_0 \approx 0.6$ (in general, $\tilde{v}_0 \sim 0.5$). Using the wave~functions of Eq. (\ref{eq:elPCFs_phi_K}) with parameters $\nu^F$ and $\tilde{k}^F_1$ ($\tilde{k}^F_0$) for edge band  $1$ ($0$) as well as the acoustic wave number $\Delta \tilde{k}^F_{0,1}$ allows for numerical evaluation of the integrals in the appendix. The interaction integrals in Eqs. (\ref{eq:F1}) and (\ref{eq:F2}) yield $F_1=-0.0546$ and $F^{\tau}_2=-0.0918$. Inserting all known values into Eq (\ref{eq:invdecay}) the resulting inverse decay time is
\begin{multline}
 \frac{1}{\tau_D} = \frac{9.1 \cdot 10^7 B[T]}{\text{s eV}^2}  \left| 0.0546 g_1 + 0.0918 g_2 \right|^2 .
\end{multline}
With the standard values\cite{PhysRevB.65.235412} of $g_1 \approx 20 \text{ eV}$ and $g_2 \approx 2 \text{ eV}$, the decay time becomes $\tau_D \approx 6.8 \text{ ns} / B[T]$, which corresponds to a characteristic decay length of $82 \text{ } \mu \text{m} / B[T]$. The decay time is much longer than the acoustic period $\sim \sqrt{B[T]} 10^{-11} \text{ s}$, thus validating our use of the Fermi golden rule.

In conclusion, we have shown that a stress-free graphene edge supports propagating vibrational in-plane edge modes in the form of 2D Rayleigh waves, and that interaction with such waves can cause electronic transitions between the electronic edge states induced by a perpendicular magnetic field. 

Since momentum conservation requires the wavelength of the acoustic waves to be on scale of the magnetic length for transitions to occur, the magnetic field strength enters into the low-amplitude absorption rate as a simple proportionality through the strain tensor. 

With the expressions given in this paper, both the absorbed acoustic frequencies and the resulting decay time can be calculated for all electronic transitions of the considered type, yielding an acoustic absorption spectrum which could be used for result confirmation in a wave-propagation experiment. 

We suggest, based on comparison with similar systems\cite{shumeiko}, that this edge-localized interaction could result in nonlinear phenomena such as acoustic solitons propagating along the edge.

\section{Acknowledgements}
We thank L. Gorelik for valuable discussion, E. Cojocaru for his Matlab scripts\cite{PCFmatlab} and the Swedish Research Council (VR) for funding. 


\bibliography{bibliography}

\begin{thebibliography}{30}
\providecommand{\natexlab}[1]{#1}
\providecommand{\url}[1]{\texttt{#1}}
\expandafter\ifx\csname urlstyle\endcsname\relax
  \providecommand{\doi}[1]{doi: #1}\else
  \providecommand{\doi}{doi: \begingroup \urlstyle{rm}\Url}\fi

\bibitem[Novoselov et~al.(2004)Novoselov, Geim, Morozov, Jiang, Zhang, Dubonos,
  Grigorieva, and Firsov]{Novoselov22102004}
K.~S. Novoselov, A.~K. Geim, S.~V. Morozov, D.~Jiang, Y.~Zhang, S.~V. Dubonos,
  I.~V. Grigorieva, and A.~A. Firsov.
\newblock Electric field effect in atomically thin carbon films.
\newblock \emph{Science}, 306\penalty0 (5696):\penalty0 666--669, 2004.
\newblock \doi{10.1126/science.1102896}.
\newblock URL \url{http://www.sciencemag.org/content/306/5696/666.abstract}.

\bibitem[Geim and Novoselov(2007)]{RiseOfGraphene}
A.~K. Geim and K.~S. Novoselov.
\newblock The rise of graphene.
\newblock \emph{Nat. mater.}, 6:\penalty0 183--191, March 2007.
\newblock \doi{10.1038/nmat1849}.
\newblock URL
  \url{http://www.nature.com/nmat/journal/v6/n3/full/nmat1849.html}.

\bibitem[Landau and Lifshitz(2008)]{ToE}
L.~D. Landau and E.~M. Lifshitz.
\newblock \emph{Theory of Elasticity}.
\newblock Butterworth-Heinemann, 2008.

\bibitem[Heil et~al.(1984)Heil, Kouroudis, Luthi, and
  Thalmeier]{0022-3719-17-13-024}
J~Heil, I~Kouroudis, B~Luthi, and P~Thalmeier.
\newblock Surface acoustic waves in metals.
\newblock \emph{Journal of Physics C: Solid State Physics}, 17\penalty0
  (13):\penalty0 2433, 1984.
\newblock URL \url{http://stacks.iop.org/0022-3719/17/i=13/a=024}.

\bibitem[Willett et~al.(1990)Willett, Paalanen, Ruel, West, Pfeiffer, and
  Bishop]{PhysRevLett.65.112}
R.~L. Willett, M.~A. Paalanen, R.~R. Ruel, K.~W. West, L.~N. Pfeiffer, and
  D.~J. Bishop.
\newblock Anomalous sound propagation at $\nu${}=1/2 in a 2d electron gas:
  Observation of a spontaneously broken translational symmetry?
\newblock \emph{Phys. Rev. Lett.}, 65:\penalty0 112--115, Jul 1990.
\newblock \doi{10.1103/PhysRevLett.65.112}.
\newblock URL \url{http://link.aps.org/doi/10.1103/PhysRevLett.65.112}.

\bibitem[Willett et~al.(1993)Willett, Ruel, Paalanen, West, and
  Pfeiffer]{PhysRevB.47.7344}
R.~L. Willett, R.~R. Ruel, M.~A. Paalanen, K.~W. West, and L.~N. Pfeiffer.
\newblock Enhanced finite-wave-vector conductivity at multiple even-denominator
  filling factors in two-dimensional electron systems.
\newblock \emph{Phys. Rev. B}, 47:\penalty0 7344--7347, Mar 1993.
\newblock \doi{10.1103/PhysRevB.47.7344}.
\newblock URL \url{http://link.aps.org/doi/10.1103/PhysRevB.47.7344}.

\bibitem[Thalmeier(2011)]{PhysRevB.83.125314}
Peter Thalmeier.
\newblock Surface phonon propagation in topological insulators.
\newblock \emph{Phys. Rev. B}, 83:\penalty0 125314, Mar 2011.
\newblock \doi{10.1103/PhysRevB.83.125314}.
\newblock URL \url{http://link.aps.org/doi/10.1103/PhysRevB.83.125314}.

\bibitem[Simon(1996)]{PhysRevB.54.13878}
Steven~H. Simon.
\newblock Coupling of surface acoustic waves to a two-dimensional electron gas.
\newblock \emph{Phys. Rev. B}, 54:\penalty0 13878--13884, Nov 1996.
\newblock \doi{10.1103/PhysRevB.54.13878}.
\newblock URL \url{http://link.aps.org/doi/10.1103/PhysRevB.54.13878}.

\bibitem[Thalmeier et~al.(2010)Thalmeier, D\'ora, and
  Ziegler]{PhysRevB.81.041409}
Peter Thalmeier, Bal\'azs D\'ora, and Klaus Ziegler.
\newblock Surface acoustic wave propagation in graphene.
\newblock \emph{Phys. Rev. B}, 81:\penalty0 041409, Jan 2010.
\newblock \doi{10.1103/PhysRevB.81.041409}.
\newblock URL \url{http://link.aps.org/doi/10.1103/PhysRevB.81.041409}.

\bibitem[Savin and Kivshar(2010)]{PhysRevB.81.165418}
Alexander~V. Savin and Yuri~S. Kivshar.
\newblock Vibrational tamm states at the edges of graphene nanoribbons.
\newblock \emph{Phys. Rev. B}, 81:\penalty0 165418, Apr 2010.
\newblock \doi{10.1103/PhysRevB.81.165418}.
\newblock URL \url{http://link.aps.org/doi/10.1103/PhysRevB.81.165418}.

\bibitem[Peres et~al.(2006)Peres, Guinea, and Castro~Neto]{PhysRevB.73.125411}
N.~M.~R. Peres, F.~Guinea, and A.~H. Castro~Neto.
\newblock Electronic properties of disordered two-dimensional carbon.
\newblock \emph{Phys. Rev. B}, 73:\penalty0 125411, Mar 2006.
\newblock \doi{10.1103/PhysRevB.73.125411}.
\newblock URL \url{http://link.aps.org/doi/10.1103/PhysRevB.73.125411}.

\bibitem[Gusynin et~al.(2008{\natexlab{a}})Gusynin, Miransky, Sharapov, and
  Shovkovy]{gusynin:778}
V.~P. Gusynin, V.~A. Miransky, S.~G. Sharapov, and I.~A. Shovkovy.
\newblock Edge states in quantum hall effect in graphene (review article).
\newblock \emph{Low Temperature Physics}, 34\penalty0 (10):\penalty0 778--789,
  2008{\natexlab{a}}.
\newblock \doi{10.1063/1.2981387}.
\newblock URL \url{http://link.aip.org/link/?LTP/34/778/1}.

\bibitem[Gusynin et~al.(2008{\natexlab{b}})Gusynin, Miransky, Sharapov, and
  Shovkovy]{PhysRevB.77.205409}
V.~P. Gusynin, V.~A. Miransky, S.~G. Sharapov, and I.~A. Shovkovy.
\newblock Edge states, mass and spin gaps, and quantum hall effect in graphene.
\newblock \emph{Phys. Rev. B}, 77:\penalty0 205409, May 2008{\natexlab{b}}.
\newblock \doi{10.1103/PhysRevB.77.205409}.
\newblock URL \url{http://link.aps.org/doi/10.1103/PhysRevB.77.205409}.

\bibitem[Gusynin et~al.(2009)Gusynin, Miransky, Sharapov, Shovkovy, and
  Wyenberg]{PhysRevB.79.115431}
V.~P. Gusynin, V.~A. Miransky, S.~G. Sharapov, I.~A. Shovkovy, and C.~M.
  Wyenberg.
\newblock Edge states on graphene ribbons in magnetic field: Interplay between
  dirac and ferromagnetic-like gaps.
\newblock \emph{Phys. Rev. B}, 79:\penalty0 115431, Mar 2009.
\newblock \doi{10.1103/PhysRevB.79.115431}.
\newblock URL \url{http://link.aps.org/doi/10.1103/PhysRevB.79.115431}.

\bibitem[Romanovsky et~al.(2011)Romanovsky, Yannouleas, and
  Landman]{PhysRevB.83.045421}
Igor Romanovsky, Constantine Yannouleas, and Uzi Landman.
\newblock Unique nature of the lowest landau level in finite graphene samples
  with zigzag edges: Dirac electrons with mixed bulk-edge character.
\newblock \emph{Phys. Rev. B}, 83:\penalty0 045421, Jan 2011.
\newblock \doi{10.1103/PhysRevB.83.045421}.
\newblock URL \url{http://link.aps.org/doi/10.1103/PhysRevB.83.045421}.

\bibitem[Castro~Neto et~al.(2009)Castro~Neto, Guinea, Peres, Novoselov, and
  Geim]{RevModPhys.81.109}
A.~H. Castro~Neto, F.~Guinea, N.~M.~R. Peres, K.~S. Novoselov, and A.~K. Geim.
\newblock The electronic properties of graphene.
\newblock \emph{Rev. Mod. Phys.}, 81:\penalty0 109--162, Jan 2009.
\newblock \doi{10.1103/RevModPhys.81.109}.
\newblock URL \url{http://link.aps.org/doi/10.1103/RevModPhys.81.109}.

\bibitem[Brey and Fertig(2006{\natexlab{a}})]{PhysRevB.73.195408}
Luis Brey and H.~A. Fertig.
\newblock Edge states and the quantized hall effect in graphene.
\newblock \emph{Phys. Rev. B}, 73:\penalty0 195408, May 2006{\natexlab{a}}.
\newblock \doi{10.1103/PhysRevB.73.195408}.
\newblock URL \url{http://link.aps.org/doi/10.1103/PhysRevB.73.195408}.

\bibitem[Castro et~al.(2010)Castro, Ochoa, Katsnelson, Gorbachev, Elias,
  Novoselov, Geim, and Guinea]{PhysRevLett.105.266601}
Eduardo~V. Castro, H.~Ochoa, M.~I. Katsnelson, R.~V. Gorbachev, D.~C. Elias,
  K.~S. Novoselov, A.~K. Geim, and F.~Guinea.
\newblock Limits on charge carrier mobility in suspended graphene due to
  flexural phonons.
\newblock \emph{Phys. Rev. Lett.}, 105:\penalty0 266601, Dec 2010.
\newblock \doi{10.1103/PhysRevLett.105.266601}.
\newblock URL \url{http://link.aps.org/doi/10.1103/PhysRevLett.105.266601}.

\bibitem[Kaasbjerg et~al.(2012)Kaasbjerg, Thygesen, and
  Jacobsen]{PhysRevB.85.165440}
Kristen Kaasbjerg, Kristian~S. Thygesen, and Karsten~W. Jacobsen.
\newblock Unraveling the acoustic electron-phonon interaction in graphene.
\newblock \emph{Phys. Rev. B}, 85:\penalty0 165440, Apr 2012.
\newblock \doi{10.1103/PhysRevB.85.165440}.
\newblock URL \url{http://link.aps.org/doi/10.1103/PhysRevB.85.165440}.

\bibitem[Nakada et~al.(1996)Nakada, Fujita, Dresselhaus, and
  Dresselhaus]{PhysRevB.54.17954}
Kyoko Nakada, Mitsutaka Fujita, Gene Dresselhaus, and Mildred~S. Dresselhaus.
\newblock Edge state in graphene ribbons: Nanometer size effect and edge shape
  dependence.
\newblock \emph{Phys. Rev. B}, 54:\penalty0 17954--17961, Dec 1996.
\newblock \doi{10.1103/PhysRevB.54.17954}.
\newblock URL \url{http://link.aps.org/doi/10.1103/PhysRevB.54.17954}.

\bibitem[Novoselov et~al.()Novoselov, Geim, Morozov, Jiang, Katsnelson,
  Grigorieva, and Dubonos]{Natv438}
K.~S. Novoselov, A.~K. Geim, S.~V. Morozov, D.~Jiang, M.~I. Katsnelson, I.~V.
  Grigorieva, and S.~V. Dubonos.
\newblock Two-dimensional gas of massless dirac fermions in graphene.

\bibitem[Jiang et~al.(2007)Jiang, Henriksen, Tung, Wang, Schwartz, Han, Kim,
  and Stormer]{PhysRevLett.98.197403}
Z.~Jiang, E.~A. Henriksen, L.~C. Tung, Y.-J. Wang, M.~E. Schwartz, M.~Y. Han,
  P.~Kim, and H.~L. Stormer.
\newblock Infrared spectroscopy of landau levels of graphene.
\newblock \emph{Phys. Rev. Lett.}, 98:\penalty0 197403, May 2007.
\newblock \doi{10.1103/PhysRevLett.98.197403}.
\newblock URL \url{http://link.aps.org/doi/10.1103/PhysRevLett.98.197403}.

\bibitem[Li et~al.(2007)Li, Andrei, and Firsov]{NatPhysv3i9}
Guohong Li, Eva~Y. Andrei, and A.~A. Firsov.
\newblock Observation of landau levels of dirac fermions in graphite.
\newblock \emph{Nat Phys}, 3:\penalty0 623, September 2007.
\newblock \doi{10.1038/nphys653}.
\newblock URL
  \url{http://www.nature.com/nphys/journal/v3/n9/suppinfo/nphys653_S1.html}.

\bibitem[Brey and Fertig(2006{\natexlab{b}})]{PhysRevB.73.235411}
L.~Brey and H.~A. Fertig.
\newblock Electronic states of graphene nanoribbons studied with the dirac
  equation.
\newblock \emph{Phys. Rev. B}, 73:\penalty0 235411, Jun 2006{\natexlab{b}}.
\newblock \doi{10.1103/PhysRevB.73.235411}.
\newblock URL \url{http://link.aps.org/doi/10.1103/PhysRevB.73.235411}.

\bibitem[Fujita et~al.(1996)Fujita, Wakabayashi, Nakada, and
  Kusakabe]{JPSJ.65.1920}
Mitsutaka Fujita, Katsunori Wakabayashi, Kyoko Nakada, and Koichi Kusakabe.
\newblock Peculiar localized state at zigzag graphite edge.
\newblock \emph{Journal of the Physical Society of Japan}, 65\penalty0
  (7):\penalty0 1920--1923, 1996.
\newblock \doi{10.1143/JPSJ.65.1920}.
\newblock URL \url{http://dx.doi.org/10.1143/JPSJ.65.1920}.

\bibitem[Arikawa et~al.(2008)Arikawa, Hatsugai, and Aoki]{PhysRevB.78.205401}
Mitsuhiro Arikawa, Yasuhiro Hatsugai, and Hideo Aoki.
\newblock Edge states in graphene in magnetic fields: A specialty of the edge
  mode embedded in the n=0 landau band.
\newblock \emph{Phys. Rev. B}, 78:\penalty0 205401, Nov 2008.
\newblock \doi{10.1103/PhysRevB.78.205401}.
\newblock URL \url{http://link.aps.org/doi/10.1103/PhysRevB.78.205401}.

\bibitem[Suzuura and Ando(2002)]{PhysRevB.65.235412}
Hidekatsu Suzuura and Tsuneya Ando.
\newblock Phonons and electron-phonon scattering in carbon nanotubes.
\newblock \emph{Phys. Rev. B}, 65:\penalty0 235412, May 2002.
\newblock \doi{10.1103/PhysRevB.65.235412}.
\newblock URL \url{http://link.aps.org/doi/10.1103/PhysRevB.65.235412}.

\bibitem[Ezawa(1971)]{ezawa}
Hiroshi Ezawa.
\newblock Phonons in a half space.
\newblock \emph{Ann. Phys.}, 67:\penalty0 438--460, 1971.

\bibitem[Borovik et~al.(1989)Borovik, Bratus, and Shumeiko]{shumeiko}
A.~E. Borovik, E.~N. Bratus, and V.~S. Shumeiko.
\newblock Hypersonic solitons in metals.
\newblock \emph{Sov. Phys. JETP}, 68:\penalty0 826--832, April 1989.

\bibitem[Cojocaru(2009)]{PCFmatlab}
E.~Cojocaru.
\newblock Parabolic cylinder functions implemented in matlab.
\newblock January 2009.
\newblock URL \url{arXiv:0901.2220}.

\end{thebibliography}
\bibliographystyle{unsrtnat} 


\appendix

\section{Electronic wave functions}
\label{sec:elwf}

The electronic pseudospinor wave~functions of the energy band $n$ and $m$, are\cite{gusynin:778,PhysRevB.77.205409,PhysRevB.79.115431,PhysRevB.83.045421}
\begin{equation}
	\vec{\psi}_{\nu}^{\tau,\tilde{k}} (x,y) = \frac{N^{\tau,k}_{\nu}}{\sqrt{L l_B}} {e}^{\imath \tilde{k} \tilde{x}} \vec{\phi}^{\tau,\tilde{k}}_{\nu} (\tilde{y}) ,
	\label{eq:elPCFs_general}
\end{equation}
where $\tau$ labels the valley ($K$ or $K'$) as before. The $y$-dependent factor is 
\begin{equation}
	\vec{\phi}^{+1,\tilde{k}}_{\nu} (\tilde{y}) =
	\begin{pmatrix}
		D_{\nu}(\sqrt{2}(\tilde{k} + \tilde{y})) \\
		\sqrt{\nu} D_{\nu-1}(\sqrt{2}(\tilde{k} + \tilde{y}))
	\end{pmatrix} \label{eq:elPCFs_phi_K} ,
\end{equation}
for the $K$-point and
\begin{equation}
	\vec{\phi}^{-1,\tilde{k}}_{\nu} (\tilde{y}) =
	\begin{pmatrix}
		\sqrt{\nu} D_{\nu -1}(\sqrt{2}(\tilde{k} + \tilde{y})) \\
		 - D_{\nu}(\sqrt{2}(\tilde{k} + \tilde{y}))
	\end{pmatrix} \label{eq:elPCFs_phi_Kpr} ,
\end{equation}
for the $K'$-point. The factors $N_{\tau,n,k}/( \sqrt{L l_B} )$ are normalization constants.

\section{Fermi level crossings}
\label{sec:fermicross}

The edge boundary condition of Eq. (\ref{eq:elbc}) ultimately gives an equation for the electronic spectrum. \emph{At the Fermi energy} $E_F$ this equation reads, for the $K$-point,
\begin{equation}
  D_{{\nu}^F} \left( \sqrt{2} \tilde{k}^F \right) = 0 , \label{eq:Kbc}
\end{equation}
and for the $K'$-point
\begin{equation}
  D_{{\nu}^F - 1} \left( \sqrt{2} \tilde{k}^F \right) = 0 , \label{eq:Kprbc}
\end{equation}
 where
\begin{equation}
  {\nu}^F = \left( \tilde{E}^F \right)^2 . \label{eq:nuF}
\end{equation}
Solving Eqs. (\ref{eq:Kbc}) and (\ref{eq:Kprbc}) for $\tilde{k}^F$ gives the Fermi crossing points $\tilde{k}^F_n$ for the given $\tilde{E}^F$. Identifying them with the different bands allows for calculation of $\Delta \tilde{k}^F_{m,n}$ and thus the absorbed acoustic frequencies. In general $\Delta \tilde{k}^F_{m,n} \sim 1$.

\section{Absorption integrals}
\label{sec:integrals}
Here the dimensionless transition integrals that enter into the transition matrix element are given. Since the integrands decay into the bulk, they are easily evaluated using a cutoff. The integral giving the scalar-potential contribution to the absorption is (normalization constants have been moved to the left hand side for brevity)
\begin{multline}
\frac{F_{1}}{N^{\tau,k^F_m  *}_{\nu^F} N^{\tau,k^F_n}_{\nu^F}} = \\
T_1
\int_{0}^{\infty} 
\vec{\phi}^{\tau,\tilde{k}^F_m \dagger}_{{\nu}^F} (\tilde{y})
\vec{\phi}^{\tau,\tilde{k}^F_n}_{{\nu}^F} (\tilde{y})
e^{- \lambda_l | \Delta \tilde{k}^F_{m,n} | \tilde{y}}
d{\tilde{y}} ,
\label{eq:F1}
\end{multline}
and the pseudo-magnetic-field contribution integral is
\begin{multline}
\frac{F_{2}}{N^{\tau,k^F_m  *}_{\nu^F} N^{\tau,k^F_n}_{\nu^F}} =
\tau
\int_{0}^{\infty} 
\vec{\phi}^{\tau,\tilde{k}^F_m \dagger}_{{\nu}^F} (\tilde{y})
\matr{\sigma}_x
\vec{\phi}^{\tau,\tilde{k}^F_n}_{{\nu}^F} (\tilde{y}) \times \\
\left( -T^{x,l}_2 e^{- \lambda_l \Delta | \tilde{k}^F_{m,n} | \tilde{y}} + T^{x,t}_2 e^{- \lambda_t | \Delta \tilde{k}^F_{m,n} | \tilde{y}}  \right)
d{\tilde{y}} + \\
\imath \sgn( \Delta \tilde{k}^F_{m,n} ) T^{y}_2 
\int_{0}^{\infty} 
\vec{\phi}^{\tau,\tilde{k}^F_m \dagger}_{{\nu}^F} (\tilde{y})
\matr{\sigma}_y
\vec{\phi}^{\tau,\tilde{k}^F_n}_{{\nu}^F} (\tilde{y}) \times \\
\left( e^{- \lambda_l \Delta | \tilde{k}^F_{m,n} | \tilde{y}} - e^{- \lambda_t | \Delta \tilde{k}^F_{m,n} | \tilde{y}}  \right)
d{\tilde{y}} .
\label{eq:F2}
\end{multline}
The numerical normalization constants are given by
\begin{equation}
\left| N^{\tau,k^F_n}_{\nu^F} \right|^2 =
\frac{1}{\int_{0}^{\infty} 
\left| \vec{\phi}^{\tau,\tilde{k}^F_n}_{{\nu}^F} (\tilde{y}) \right|^2
d{\tilde{y}}} .
\label{eq:normN}
\end{equation}

\end{document}